\documentclass[journal=jctcce,manuscript=article,layout=twocolumn]{achemso}
\usepackage{graphicx}
\usepackage{amsmath}
\usepackage{xcolor}
\usepackage{amssymb}
\usepackage{enumerate}
\def\br{{\bf r}}
\def\bp{{\bf r^\prime}}

\def\dd{{\rm d}}

\date{}

\title{Stochastic vertex corrections: linear scaling methods for accurate quasiparticle energies}
\author{Vojt\v{e}ch Vl\v{c}ek}
\affiliation{Department of Chemistry and Biochemistry, University of California, Santa Barbara, 93106, U.S.A}
\email{vlcek@ucsb.edu}

\begin{document}
\maketitle

\begin{abstract}
New stochastic approaches  for the computation of electronic excitations are developed within the many-body perturbation theory. Three approximations to the electronic self-energy are considered: $G_0W_0$, $G_0W_0^tc$, and $G_0W_0^{tc}\Gamma_x$. All three methods are formulated in the time domain and the latter two incorporate non-local vertex corrections.  In case of $G_0W_0^{tc}\Gamma_x$, the vertex corrections are included both in the screened Coulomb interaction and in the expression for the self-energy. The implementation of the three approximations is verified by comparison to deterministic results for a set of small molecules. The performance fully stochastic implementation is tested on acene molecules, C$_{60}$ and PC$_{60}$BM.  The vertex correction appears crucial for the description of unoccupied states. Unlike conventional (deterministic) approaches, all three stochastic methods scale linearly with the number of electrons.
\end{abstract}

\section{Introduction}

Efficient first-principles methods allow calculations of the ground state electronic structure in large molecules and solids. \cite{goedecker1999linear,goedecker2003linear,skylaris2005introducing,vandevondele2005quickstep,zhou2006self,neese2009efficient,saad2010numerical,baer2013self,neuhauser_comm_2014} However, quantitative prediction of electronic excitation is still computationally prohibitive. Traditional quantum chemistry methods, such as configurational interaction\cite{ostlund1996modern,sherrill1999configuration} or coupled cluster\cite{cizek1966correlation,paldus1972correlation,bartlett2007coupled,krylov2008equation} approaches scale at least as $N_e^6$ (where $N_e$ is the number of electrons).\cite{goedecker2003linear,kaliman2017new} As a result, these methods are applied only to small systems. 

Many-body perturbation theory\cite{fetter2003quantum,martin2016interacting} offers an alternative and becomes an increasingly popular tool for computation of quasiparticle (QPs) energies of molecules. The central quantity is the QP self-energy, i.e.,  a dynamical potential that embodies all many-body interactions. In principle,  it is found in a self-consistent manner.\cite{hedin1965new,martin2016interacting} The self-consistency relates the self-energy to the QP Green's function, polarizability and screened Coulomb interaction. 

In practice, the expression for the self-energy is often simplified by neglecting high-order terms (non-trivial part of vertex function $\Gamma$) leading to the $GW$ approximation.\cite{hedin1965new,aryasetiawan1998gw,onida2002electronic,martin2016interacting} In addition, self-consistency in $GW$ calculations is either further approximated or completely avoided.\cite{hybertsen1985first,hybertsen1986electron,aryasetiawan1998gw,faleev2004all,van2006quasiparticle,stan2009levels,bruneval2014quasiparticle,martin2016interacting}  The latter thus corresponds to a one-shot correction, labeled $G_0W_0$, on top a mean-field starting point (usually DFT). Such practical simplification of $GW$ has two purposes: (i) Conventional implementations scale as $N_e^3$ or $N_e^4$, and repeated evaluation of the self-energy is  thus costly even for small systems.\cite{deslippe2012berkeleygw,nguyen2012improving,pham2013g,liu2016cubic} (ii) Self-consistent $GW$ may yield worse results than $G_0W_0$ due to the absence of the vertex term.\cite{holm1998fully,bruneval2014quasiparticle,cao2017fully} The typical strategy is thus to use $G_0W_0$ on top of the ``best'' possible DFT starting point.\cite{bruneval2012benchmarking,marom2012benchmark,knight2016accurate,caruso2016benchmark,rangel2016evaluating} Recent benchmark for acenes, however, revealed that $GW$ suffers from substantial errors for QP energies of unoccupied states.\cite{rangel2016evaluating}

Beyond $GW$ techniques include approximate vertex functions ($\Gamma$), which are closely related to the electron-hole interaction kernel in the Bethe-Salpeter equation (BSE).\cite{onida2002electronic} In practice $\Gamma$ is computed in various ways: Local vertex functions derived from the Kohn-Sham time-dependent density functional theory (TDDFT) are simple and relatively inexpensive,\cite{del1994gwg,hung2016excitation,hellgren2018beyond,hellgren2018local,ma2018finite} but they do not remedy failures of $GW$, such as spurious ``self-screening error''.\cite{romaniello2009self} Furthermore, they do not outperform simple $G_0W_0$.\cite{morris2007vertex,hung2016excitation,ma2018finite}  

Non-local vertex functions seem to improve the description of the QP energies;\cite{shishkin2007accurate,romaniello2009self,ayral2012spectral,kuwahara2016gw,maggio2017gw} however, they are costly and suffer from steep scaling ($N_e^6$ ).\cite{maggio2017gw}  Alternatively, the vertex term has been approximated up to the second order,\cite{freeman1977coupled,ren2015beyond,knight2016accurate} but this is associated with only mild cost reduction ($N_e^5$ scaling).\cite{gruneis2009making,ren2015beyond} Consequently, the beyond $GW$ calculations have been applied only to model or few-electron systems.\cite{shishkin2007accurate,romaniello2012beyond,kuwahara2016gw,maggio2017gw}

Here,  numerical and theoretical developments are combined to overcome this limitation. A self-consistent expression for the self-energy with non-local $\Gamma$ is constructed using derivatives of the inverse Green's function.\cite{romaniello2012beyond,starke2012self}  In practice, we apply only a one-shot correction, in which $\Gamma$ is derived from the non-local exchange term present in the mean-field starting point: either in the Hartree-Fock (HF) approximation or generalized Kohn-Sham (GKS) theory.\cite{Seidl1996}  The approach is labeled $GW\Gamma_X$. 

To lower the computational cost, $GW\Gamma_X$ is implemented using real-time \emph{stochastic} numerical techniques.\cite{baer2013self,neuhauser2014breaking,gao2015sublinear,neuhauser2015stochastic,rabani2015time,vlcek2017stochastic,vlvcek2018swift} Up to now, stochastic calculations of QP energies were limited to $G_0W_0$ with DFT based on the local density approximation (LDA) to exchange and correlation (xc). Here, we first extend the methodology to HF and hybrid xc functionals. Next, the stochastic form of the self-energy is presented and tested on a set of molecules. 

The stochastic implementation scales (sub)linearly with the number of electrons. Furthermore, favorable self-averaging leads to low statistical noise. For large systems, the $GW\Gamma_X$ method is found to be computationally less expensive than the stochastic implementation of $G_0W_0$ based on hybrid xc functionals. Results for ionization potentials and electron affinities of molecules suggest that the inclusion of a non-local vertex is necessary for accurate predictions of QP energies.

The manuscript is organized as follows: Derivation of the self-energy expressions is presented in Sec.~\ref{sec:theory}. The stochastic formulation suitable for numerical implementation is shown in Sec.~\ref{sec:methods}. Performance of the method and its implementation are demonstrated in the results section (Sec.~\ref{sec:results}) followed by conclusions and outlook  (Sec.~\ref{sec:conclusions}).

\section{Theory}\label{sec:theory}

In this section, we first review the theoretical description of quasiparticles (QP), namely quasi-electron and quasi-holes. Propagation of a QP is described by a single-particle Green's functions (GF), which is defined as 
\begin{equation}
iG(1,2) = \left\langle \Psi \middle | \mathcal{T}\, \hat\psi(1) \hat\psi^\dagger(2) \middle | \Psi \right\rangle,
\end{equation}
where $\Psi$ is the ground state many-body wave-function of the $N_e$-electron system, $\mathcal{T}$ denotes time ordering operator, $\hat\psi$ and $\hat\psi^\dagger$ are the electron annihilation and creation operators. Here, we adopt a short-hand notation for space-time coordinates: $\left(\br_1, t_1\right) \equiv 1$. 

The GF satisfies the equation of motion 
\begin{equation}
\left[i\frac{\partial}{\partial t} - \hat{h}\right] G(1,2) - \Sigma_T(1,\bar3) G(\bar3,2)  = \delta(\br_1-\br_2)\label{GFeom}
\end{equation}
where  $\hat{h}$ contains the kinetic energy and the electron-nuclear attraction terms, and $\Sigma_T$ is the \emph{total} self-energy, which contains both Hartree and exchange-correlation interactions. Further, we simplified the notation by omitting integration symbol and introducing a bar symbol above the space-time coordinates that should be integrated. 

In this work, we focus on QP energies of quasi-electrons and quasi-holes  ($\varepsilon$), which are obtained from the QP equation:
\begin{equation}
\hat{h} \psi(\br_1) + \int \Sigma_T(\br_1, \br_2, \omega=\varepsilon) \psi(\br_2) \dd \br_2 = \varepsilon \psi(\br_1)\label{QPequation}
\end{equation}
where $\psi$ is the QP  state. Eq.~\ref{QPequation} is a fixed point expression where $\Sigma_T$ has to be computed at the frequency corresponding to $\varepsilon$. In the rest of the paper, we use a real-time representation of the self-energy, $\Sigma_T(\br_1, \br_2,t_1,t_2)$ which is merely a Fourier transformation of $\Sigma_T(\br_1, \br_2,\omega)$. 

We will now review the expressions for the total self-energy in Sec.~\ref{subsec:sigma_general} and then we turn to the approximate forms in Sec.~\ref{subsec:sigma_approx}.

\subsection{Self-energy}\label{subsec:sigma_general}

The total self-energy ($\Sigma_T$) in principle requires knowledge of the two-particle GF, leading to a hierarchy of coupled equations of motion.\cite{romaniello2012beyond,starke2012self,martin2016interacting} Alternatively, $\Sigma_T$ is written as a sum:\cite{hedin1965new,aryasetiawan1998gw}
\begin{equation}
\Sigma_T (1,2) =  \Sigma_H (1)\delta(1,2) +   \Sigma_{xc}, (1,2), \label{sigma_total}
\end{equation}
where $\Sigma_H$ and $\Sigma_{xc}$ are the Hartree and exchange-correlation self-energies. Note that the former is local and instantaneous; hence, it appears together with a delta function $\delta(1,2) \equiv \delta(\br_1 - \br_2) \delta(t_1 - t_2)$.

The Hartree term represents the interaction with the electron density:
\begin{equation}
\Sigma_H (1) = -i \nu(1,\bar 2) G(\bar 2,\bar 2^+)\label{sigma_H}
\end{equation}
where $\nu$ is the instantaneous Coulomb interaction defined as
\begin{equation}
\nu(1,2) = \frac{1}{\left|\br_1 - \br_2\right|} \delta(t_1 - t_2)\label{coulomb_kernel}
\end{equation}
and the density is given by the equal-time GF, i.e., $n(\br_1) \equiv G(1,1^+)$. The $1^+$ argument represents $(\br_1,t_1^+)$, where $t_1^+$ is only infinitesimally after $t_1$. 

The exchange-correlation self-energy is:
\begin{equation}
\Sigma_{xc}(1,2) =  -i\nu(1,\bar{4})G(1,\bar{3}) \frac{\delta G^{-1} (\bar{3},2)}{\delta U(\bar{4})}, \label{sigma_xc_invG}
\end{equation}
where $U$ is  an external potential introduced to remove the two-particle GF in the expression for $\Sigma_T$.\cite{hedin1965new,aryasetiawan1998gw}  
The derivative of the inverse GF in Eq.~\ref{sigma_xc_invG} leads to two equivalent expressions: The first one is very compact and includes a three-point \emph{irreducible vertex function} $\Gamma$; the second is slightly more involved, but it is more versatile expression leading to useful approximations to $\Gamma$. For completeness, we will now review both. 

\subsubsection{ $\Sigma_{xc}$ with the vertex function}
In the first route, we consider a chain rule of derivatives:
\begin{equation}
\frac{\delta G^{-1} ({3},2)}{\delta U({4})} = \frac{\delta G^{-1} ({3},2)}{\delta U_{cl}(\bar{5})}\frac{\delta U_{cl}(\bar{5})}{\delta U({4})},\label{chain_invGU}
\end{equation}
where $U_{cl}$ is a classical potential, consisting of the Coulomb and external potentials. From classical electrostatics, the change of $U_{cl}$ with the variation of the external potential $U$ corresponds to the inverse dielectric function:
\begin{equation}
\epsilon^{-1}({5},{4}) \equiv \frac{\delta U_{cl}({5})}{\delta U({4})}.\label{inv_diel_fc1}
\end{equation}
The first derivative in the right side of Eq.~\ref{chain_invGU} serves as the definition of the irreducible vertex function:
\begin{equation}
\Gamma({3},2,{5}) := -\frac{\delta G^{-1} ({3},2)}{\delta U_{cl}({5})}.\label{def_ir_gamma}
\end{equation}
Combining Eqs.~\ref{sigma_xc_invG},\ref{inv_diel_fc1} and \ref{def_ir_gamma} leads to the following compact expression for the exchange-correlation self-energy:
\begin{equation}
\Sigma_{xc}(1,2) =  iW(1,\bar{4})G(1,\bar{3}) \Gamma(\bar{4},\bar{3}, 2), \label{sigma_GWg}
\end{equation}
where $W$ is the screened Coulomb interaction obtained by convolution of Eqs.~\ref{coulomb_kernel} and \ref{inv_diel_fc1}.

\subsubsection{$\Sigma_{xc}$  with generalized polarizability}
In the second route, we start again from Eq.~\ref{sigma_xc_invG} and make use of the functional derivatives.\cite{romaniello2012beyond,starke2012self} First, the change of the inverse GF with respect to $U$ is:
\begin{align}
&\Sigma_{xc}(1,2) =\nonumber \\ 
&i\nu(1,\bar{4})G(1,\bar{3}) \left[ \delta(\bar{3},2)\delta(\bar{3},\bar{4}) + \frac{\delta \Sigma_T (\bar{3},2)}{\delta U(\bar{4})} \right].\label{sigma_xc_scf}
\end{align}
The two terms in the square brackets lead to a suitable definition of the exchange and correlation self-energies. The former is
\begin{equation}
\Sigma_x (1,2) = i \nu(1,2) G(1,2)\label{sigma_x}.
\end{equation}
Note that the Coulomb kernel is instantaneous (Eq.~\ref{coulomb_kernel}), so $G(1,2)$ in Eq.~\ref{sigma_x} is the density matrix $\rho(1,2)\equiv G(1,2) \delta(t_1-t_2) $. 

The correlation self-energy has a more complicated expression:
\begin{equation}
\Sigma_c (1,2) = i\nu(1,\bar{4})G(1,\bar{3})  \frac{\delta \Sigma_T (\bar{3},2)}{\delta U(\bar{4})} .\label{sigma_P}
\end{equation}

It is convenient to recast the functional derivative of the total self-energy in Eq.~\ref{sigma_P} as
\begin{equation}
\frac{\delta \Sigma_T ({3},2)}{\delta U({4})}  =  \frac{\delta \Sigma_T ({3},2)}{\delta G (\bar{6},\bar{5})}  \frac{\delta G (\bar{6},\bar{5})}{\delta U({4})}.
\end{equation}
Further, we introduce a generalized three-point reducible polarizability:
\begin{equation}
^3\chi({6},{5}, {4}) :=  -\frac{\delta G ({6},{5})}{\delta U({4})}.\label{3chipol}
\end{equation}
The final expression of the correlation self-energy thus reads
\begin{equation}
\Sigma_c (1,2) =- i\nu(1,\bar{4})G(1,\bar{3})  \frac{\delta \Sigma_T (\bar{3},2)}{\delta G (\bar{6},\bar{5})}  ~^3\chi(\bar{6},\bar{5}, \bar{4}) .\label{sigma_P_2}
\end{equation}

The compact expression with the vertex function (Eq.~\ref{sigma_GWg}) is equivalent to the sum of the exchange (Eq.~\ref{sigma_x}) and polarization (Eq.~\ref{sigma_P_2}) self-energies. Note that $\Sigma_c$ depends on a functional derivative of $\Sigma_T$ (Eq.~\ref{sigma_P_2});  hence, the total self-energy should be, in principle, found by a self-consistent cycle.

\subsection{Approximate Self-energy}\label{subsec:sigma_approx}

Due to the quadruple integration, Eq.~\ref{sigma_P_2} is computationally difficult. In the following, we outline how to construct practical expressions for $\Sigma_c$ based on successive approximations for the functional derivative $\delta \Sigma_T/ \delta G$.

\subsubsection{The $GW$ approximation}

In the $GW$ approximation, the total-self-energy in Eq.~\ref{sigma_P_2} is substituted with the classical Hartree self-energy $\Sigma_H$. Hence, the functional derivative of $\Sigma_T$ with respect to the GF becomes
\begin{equation}
 \frac{\delta \Sigma_T ({3},2)}{\delta G ({6},{5})}  \approx  -\nu(2,{5}) \delta({6},{5})\delta({3},2),\label{DSt_dG_GW}
\end{equation}
where the delta functions are due to the locality of $\Sigma_H$ (Eq.~\ref{sigma_H}). 

The approximation introduced in Eq.~\ref{DSt_dG_GW} greatly simplifies the polarization self-energy: $\, ^3\chi$  becomes two-point reducible polarizability, i.e., 
\begin{equation}
\chi({5}, {4}) := - \frac{\delta G ({5},{5}^+)}{\delta U({4})} = \frac{\delta n(5)}{\delta U(4)}.\label{polarizability_2pt}
\end{equation}
Note the $\chi$ is a time-ordered quantity, but it is trivially related to the standard retarded response function.\cite{vlcek2017stochastic,vlvcek2018swift}
Consequently, the polarization self-energy has the following form:
\begin{equation}
\Sigma^{GW}_c (1,2) = i\nu(1,\bar{4})G(1,2) \nu(2,\bar5)  \chi(\bar{5}, \bar{4}).\label{sigmaP_gw}
\end{equation}

The $GW$ xc self-energy is a sum of Eqs.~\ref{sigma_x} and~\ref{sigmaP_gw}, which becomes 
\begin{equation}
\Sigma_{xc}^{GW} (1,2) =  iW(1,2^+)G(1,2), \label{GWapprox}
\end{equation}
where we used an alternative definition of the screened Coulomb interaction:
\begin{equation}
W(1,2) = \nu(1,2) + \nu(1,\bar{3}) \chi(\bar3,\bar4) \nu(\bar4,2).
\end{equation}

Note that the $GW$ self-energy is often obtained from Eq.~\ref{sigma_GWg} by approximating the vertex function as $\Gamma(1,2,3)\approx \delta(1,2)\delta(1,3)$. Such derivation of $\Sigma_{xc}^{GW} $  is quite simple and compact, but it is not immediately clear how to construct a better approximation. 

\subsubsection{$GW\Gamma_X$ approximation}

In this part, we will consider the next step in the construction of the self-energy. We use Eq.~\ref{sigma_P_2},  and take $\Sigma_T\approx \Sigma_H + \Sigma_x$. Note that this expression omits the functional derivative of the correlation self-energy.  We denote this approximation $GW\Gamma_X$. The derivative of the total self-energy becomes:
\begin{align}
&\frac{\delta \Sigma_T ({3},2)}{\delta G ({6},{5})}  \approx\nonumber \\ &- \nu(2,{5}) \delta({6},{5})\delta({3},2) + \nu(3,2) \delta({5},2) \delta(6,{3}) \label{DSt_dG_GWG}
\end{align}
The first term on the right is the classical Hartree interaction (as in Eq.~\ref{DSt_dG_GW}), the second term is due to a non-local exchange. While both terms include the Coulomb kernel (Eq.~\ref{coulomb_kernel}),  they have a different structure, i.e., each contracts distinct space-time points.  As a result, the polarization self-energy contains a contribution from the reducible three-point polarizability:
\begin{align}
\Sigma^{GW\Gamma_X}_c (1,2) =& i\nu(1,\bar{4})G(1,2)  \nu(2,\bar5)  \chi(\bar{5}, \bar{4}) \nonumber \\
 -&i\nu(1,\bar{4})G(1,\bar3)  \nu(\bar3,2)  \,^3\chi(\bar{3},2, \bar{4}).\label{sigmaP_gwg}
\end{align}
The definition of $\nu$ (Eq.~\ref{coulomb_kernel}) contains a delta function which guarantees that the Coulomb interaction is instantaneous in time. Hence, the second term in Eq.~\ref{sigmaP_gwg} implicitly contains $\delta(t_{2}-t_{\bar3})$.   As a result, the generalized polarizability depends on three spatial coordinates but only two time points. In other words, $\,^3\chi (\bar{3},2, \bar{4})$ (defined by Eq.~\ref{3chipol}) yields time-dependent induced density matrix $\delta\rho(\br_{\bar3},\br_2, t_{\bar 3},t_2) \delta(t_2-t_{\bar3})$ due to the variation of the external potential at $4$.  The response of the density matrix is time-ordered with respect to $t_4$.  In practice, $\,^3\chi (\bar{3},2, \bar{4})$  is not computed explicitly;  as shown in Sec.~\ref{sssec:stochGFSE},  the second term  is evaluated using real-time propagation of the time-dependent induced density matrix.

Finally, it is important to comment on the relation between $GW\Gamma_X$ approximation and the second-order screened-exchange (SOSEX) method.\cite{ren2013renormalized,ren2015beyond,knight2016accurate}  In the latter, two distinct steps are involved in approximating Eq.~\ref{sigma_P_2}: (i) $\Sigma_T\approx \Sigma_H + \Sigma^{GW}_{xc}$, and the functional derivative of the second term is $\delta \Sigma^{GW}_{xc}/ \delta G \approx W$   (ii) the three-point polarizability is $\,^3\chi (1,2,3) \approx G(1,3) G(3,2)$, which can be viewed as a generalized case of the independent QP approximation.\cite{onida2002electronic} 

Together, the  $GW$+SOSEX self-energy is\cite{ren2013renormalized,ren2015beyond}
\begin{align}
&\Sigma^{GW+SOSEX}_c (1,2) = i\nu(1,\bar{4})G(1,2)  \nu(2,\bar5)  \chi(\bar{5}, \bar{4}) \nonumber \\
 -&i\nu(1,\bar{4})G(1,\bar3) G(\bar3,\bar4)G(\bar4,2) W(\bar3,2) .
\end{align}
Clearly, this expression is different from Eq.~\ref{sigmaP_gwg}. Unlike $GW\Gamma_X$, it contains a screened Coulomb interaction, and it lacks the induced density matrix. SOSEX approximates the vertex to second order\cite{maggio2017gw} and, as discussed,  it represents a distinct approach to solve Eq.~\ref{sigma_P_2}. 

In the rest of the paper, we will consider only $GW$ (with and without RPA - c.f., Sec.~\ref{sec:tprop_stoch_LRX}) and the  $GW\Gamma_X$ self-energies, in which the vertex term is derived from the mean-field starting-point.  

\section{Computational methodology}\label{sec:methods}

In this section, we present practical steps which allow the application of Eqs.~\ref{sigmaP_gw} and \ref{sigmaP_gwg} to large molecules. In practice, we employ two simplifications: 

(i) We do not seek self-consistency in $\Sigma_T$. The $i^{\rm th}$ QP energy is computed as: 
\begin{equation}
\varepsilon_i = \varepsilon_i^{0} + \left \langle \phi_i \middle | \hat{\Sigma}(\omega = \varepsilon_i) - \hat{v}_{xc} \middle | \phi_i \right \rangle,\label{QP_eq_lin_DFT}
\end{equation}
where $\varepsilon_i^{0}$  is an eigenvalue of the mean-field Hamiltonian, $\phi_i$ is the corresponding eigenstate, and $\hat{v}_{xc}$ is the mean-field exchange-correlation potential operator. 
The one-shot correction means that the GF and the screened Coulomb interaction (denoted $G_0$ and $W_0$) are expressed using the mean-field eigenvalues and eigenstates. 
Further, $\,^3\chi$ and $\chi$ in Eq.~\ref{sigmaP_gwg} are substituted with  $\,^3\chi_0$ and  $\chi_0$, which correspond to the response functions computed with the mean-field Hamiltonian

In a one-shot correction scheme, $\delta \Sigma_x / \delta G_0$ is obtained from the derivative of the mean-field non-local exchange. Here, we consider HF and GKS starting points.  In the second case, only the \emph{non-local} part of the exchange (introduced in Eq.~\ref{sigma_gx} in Sec.~\ref{ssec:RSH_DFT}) is considered.

(ii) We reformulate Eqs.~\ref{GWapprox} and \ref{sigmaP_gwg} using the stochastic approach, i.e., the expectation values become  statistical estimators over (many) stochastic samples. This method was applied to the $G_0W_0$ approximation as described in earlier publications.\cite{neuhauser2014breaking,vlcek2017stochastic,vlvcek2018swift} In contrast to the previous work, the starting point is no longer constrained to DFT with local functionals. The stochastic formulation of $\Sigma_c^{GW\Gamma_X}$ is a new development.

The details of the starting point Hamiltonians are given in the next subsection (\ref{ssec:RSH_DFT}), followed by a short overview of the stochastic approach and the description of the new developments (\ref{ssec:stochastics}).

\subsection{Mean-field starting points}\label{ssec:RSH_DFT}

The starting point is computed with a Hamiltonian:
\begin{equation}
\hat{H}_{0} = \hat{h} + \hat\Sigma_H  + \hat{V}_c + \hat{V}_x + \hat\Sigma_x^\gamma\label{RSHHamiltonian}
\end{equation}
where $\hat{h}$ contains the kinetic energy and the electron-nuclear attraction (as in Eq.~\ref{GFeom}),  and $V_c$ is the correlation potential approximated by a semilocal functional of the density. The exchange interaction is based on spatial separation of the $1/r$ kernel into short and long-range parts.\cite{Savin1995,Leininger1997,BaerNeuhauser2005}   $\Sigma_x^\gamma$ is the non-local long-range exchange interaction \begin{equation}
\Sigma_x^\gamma (1,2) = i \nu^\gamma(1,2) G_0(1,2)\label{sigma_gx},
\end{equation}
where $G_0$ is the GF constructed from the $H_0$ and $\nu^\gamma$ is the exchange kernel:
\begin{equation}
\nu^\gamma(1,2) = \frac{{\rm erf}\left(\gamma r\right)}{\left|\br_1 - \br_2\right|} \delta(t_1 - t_2).\label{coulomb_g_kernel}
\end{equation}
The short-range part, $V_x$, is derived from the complementary error-function term; it is given by a semilocal density functional which depends on the value of $\gamma$.
In HF calculations, the $V_c$ and $V_x$  are set to zero, and the non-local exchange is given by Eq.~\ref{sigma_x} (i.e., the range-separation parameter $\gamma\to\infty$).

The calculations presented in this work rely on two starting points: HF and the optimally tuned LC-$\omega$PBE functional\cite{vydrov2006assessment} implemented using the LibXC library.\cite{marques2012libxc,lehtola2018recent} In practice, optimal tuning amounts to finding range-separation parameter $\gamma$ which enforces the IP theorem, i.e., $\gamma$ is varied such that the negative of the HOMO energy corresponds to the ionization potential (energy difference between a neutral system and a cation).  Optimal tuning is associated with mitigation of spurious electron self-interaction and leads to good $I$ and fundamental band gaps ($E_g$) in finite systems \cite{Stein2010,Stein2012,Kronik2012}.  Further, TDDFT with optimally tuned functionals with long-range exchange treats attractive electron-hole interactions and efficiently mimics BSE.\cite{yang2015simple,refaely2015solid,brawand2016generalization}

\subsection{Stochastic approach}\label{ssec:stochastics}

\subsubsection{Stochastic sampling of Green's functions and self-energies}\label{sssec:stochGFSE}
We will now introduce the basics of the stochastic approach and describe how $\Sigma^{GW\Gamma_X}$ and $\Sigma^{GW}$ are computed using stochastic sampling of the GFs.

In the initial part of the algorithm, random functions are prepared on a real space grid as: 
\begin{equation}
\bar\zeta(\br) = \pm \frac{1}{\sqrt{\dd \Omega}}
\end{equation}
where $\dd \Omega$ is the volume element associated with each grid point. The $\pm$ in front of the fraction represents a random sign assigned to each space-point $\br$. This choice satisfies the stochastic resolution of identity $\hat{\mathcal I} \equiv \left\{ \left| \bar\zeta\middle\rangle \middle \langle \bar \zeta\right| \right\}$, where $\hat{\mathcal{I}}$ is the identity operator and $\left\{\cdots\right\}$ denotes an average over the entire (in principle infinite) set of random functions. 

In the stochastic representation, the density matrix is given as an average:
\begin{equation}
\rho\left(\br_1,\br_2\right) = \left\{ \eta \left(\br_1\right)  \eta^* \left(\br_2\right) \right\}.\label{stodensmat}
\end{equation}
Here, $\eta$ are random vectors within the \emph{occupied} subspace, i.e., $\left| \eta \right\rangle = \hat{P} \left| \bar\zeta \right\rangle$ and $\hat{P}$ is a projection operator.  $\hat{P}$ depends on the chemical potential and the Hamiltonian $\hat{H}_0$. In practice, $\eta$ states can be constructed either by projecting on the occupied eigenstates or, e.g., by Chebyshev filtering.\cite{baer1997chebyshev,baer2013self,neuhauser2014breaking,vlcek2017stochastic,vlvcek2018swift} In this paper, we follow the former approach.

The stochastic form of the GKS Green's function $G_0$ is\cite{neuhauser2014breaking,vlcek2017stochastic,vlvcek2018swift}
\begin{equation}
iG_0(\br_1,\br_2,t_1-t_2)=\left\{ \zeta(\br_1,t_1)\bar{\zeta}(\br_2,t_2) \right\},\label{Gtexpansion}
\end{equation}
where the $\zeta$ vector is either in the occupied or unoccupied subspace, i.e., it is obtained by projection with $\hat{P}$ or its complement $\left( \hat{\mathcal{I}} - \hat{P}\right)$. Since the  equilibrium GF depends only on the difference between $t_1$ and $t_2$, we set $t_1=0$ and let only the projected stochastic vectors to evolve in time. For negative/positive times, the GF represents a propagator of holes/electrons. The corresponding time-evolved random vectors are:
\begin{equation}
\zeta(\br,t) = \begin{cases} \left\langle \br \middle |e^{-i\hat{H}_0t}\hat{P}\middle |\bar\zeta\right\rangle & t <0 \\ \left\langle \br \middle |e^{-i\hat{H}_0t}\middle(\hat{\mathcal{I}}-\hat{P}\middle)\middle |\bar\zeta\right\rangle & t >0.\end{cases}\label{tprop_zeta}
\end{equation}
In practice, the time propagation is performed using Trotter (split operator) technique. The computational cost of the time propagation scales with the number of occupied states in $\hat H_0$. It is possible to reduce the cost by employing stochastic time propagation described in Sec.~\ref{sec:tprop_stoch_LRX} and in Refs.~\citenum{neuhauser2014breaking,vlcek2017stochastic,vlvcek2018swift,vlcek2019stochastic}.

We will now focus on the two approximations introduced earlier (Eqs.~\ref{sigmaP_gw} and \ref{sigmaP_gwg}) and combine them with the stochastic form of the GF (Eq.~\ref{Gtexpansion}). 

The $G_0W_0$ approximation to $\Sigma_c$ is:\cite{neuhauser2014breaking,vlcek2017stochastic,vlvcek2018swift}
\begin{equation}
\Sigma^{GW}_c (t) = \big\{ \phi_i(\bar\br_1) \zeta (\bar\br_1,t) W_P(\bar\br_1, t) \big\}, \label{sto_sigmaP_gw_deln}
\end{equation}
where we introduced a time-ordered polarization potential $W_P(t) = \hat{\nu}\, \hat{\chi}_0(t) \,\hat{\nu} \left| \bar\zeta \phi_i\right\rangle$, which is obtained from a retarded potential, $W_P^r$, by manipulating its imaginary components in the frequency domain\cite{neuhauser2014breaking,vlcek2017stochastic,vlvcek2018swift} using sparse stochastic compression technique\footnote{Details of the implementation are given in Vlcek et al. Phys. Rev. B, 98(7):075107, 2018. Here, we use 20,000 stochastic vectors with size of 10\% of the total real-space grid. This was found to be sufficient and agrees with our previous finding both molecules and periodic systems.}. Note that the response function $\chi_0$ is based on the mean-field starting point (i.e., HF or GKS) as discussed earlier.

The retarded potential is computed as 
\begin{equation}
W^r_P(\br, t) = {\nu}(\br,\bar\br_2) \delta{n}(\bar\br_2, t) 
\end{equation}
where the induced time-dependent density is
\begin{equation}
\delta{n}(\br, t) = {\chi}_0(\br,\bar\br_2,t) {\nu}(\bar\br_2,\bar\br_3)\bar\zeta(\bar\br_3) \phi_i(\bar\br_3).\label{deltan_response_pot}
\end{equation}
In practice, the induced density is computed as a difference $\delta{n}(\br, t) = {n}(\br, t)- {n}(\br,0)$ , where the density is constructed from time-evolved occupied states ${n}(\br, t) = \sum_i^{N_{occ}} |\phi(\br,t)|^2$.  The system is perturbed at $t=0$ by  a potential  ${\nu}(\br,\bar\br_2)\bar\zeta(\bar\br_2) \phi_i(\bar\br_2)$. 

In the fully stochastic formulation, $\delta{n}(\br, t)$ is computed via stochastic sampling detailed in Sec.~\ref{sssec:stochdenmtx}. Further, the propagation is performed using: (i) random phase approximation (RPA) or (ii) TDDFT. Both approaches are discussed at the end of Sec.~\ref{sec:tprop_stoch_LRX}.

Eq.~\ref{sto_sigmaP_gw_deln} provides an intuitive interpretation of the $GW$ correlation self-energy: it is a time-dependent induced Coulomb potential due to the addition of an electron/hole to the state $\phi$.

The correlation self-energy in the $GW\Gamma_X$ approximation is based on Eq.~\ref{sigmaP_gwg}. Using the expression for the $GW$ self-energy, we obtain
\begin{align}
\Sigma^{GW\Gamma_X}_c (t) =&  \left\{ \phi_i(\bar\br_1) \zeta (\bar\br_1,t) W_P(\bar\br_1, t) \right. \nonumber \\ &+ \left.\phi_i(\bar\br_1) \zeta (\bar\br_2,t)  W_x(\bar\br_1, \bar\br_2 , t)  \right\},\label{sto_sigmaP_gwg_deln}
\end{align}
where we introduced a time-ordered induced exchange potential $W_x(t) = \hat{\nu}^\gamma\, ^3\hat{\chi}_0(t) \,\hat{\nu} \left| \bar\zeta \phi_i\right\rangle$, which contains the exchange kernel $\nu^\gamma$ (Eq.~\ref{coulomb_g_kernel}) and the three-point polarizability $^3\hat{\chi}_0(t)$. The polarizability is based on the mean-field starting point (i.e., HF or GKS). The induced exchange potential is computed from its retarded form $W^r_x(\br, t) $. The time-ordering is performed in the frequency domain and $W_P$ acquires positive/negative sign for electrons/holes. 

The retarded potential is computed as:
\begin{equation}
W^r_x(\br, \bp, t) = {\nu}^\gamma(\br,\bp) \delta{\rho}(\br,\bp, t), 
\end{equation}
where the induced time-dependent density matrix is
\begin{equation}
\delta{\rho}(\br,\bp, t) = \,^3{\chi}_0(\br,\bp,\bar\br_2,t) {\nu}(\bar\br_2,\bar\br_3)\bar\zeta(\bar\br_3) \phi_i(\bar\br_3).
\end{equation}
In practice, it is computed as a difference $\delta{\rho}(\br,\bp, t) = {\rho}(\br,\bp, t)- {\rho}(\br,\bp,0)$ , the density matrix is constructed from time-evolved occupied states ${\rho}(\br,\bp, t) = \sum_i^{N_{occ}} \phi(\br,t) \phi^*(\bp,t)$. In the fully stochastic formulation, $\delta{\rho}(\br,\bp, t)$ is computed via stochastic sampling detailed in Sec.~\ref{sssec:stochdenmtx}.
Hence, $\Sigma_c^{GW\Gamma_X}$ contains time-dependent induced Coulomb and exchange potentials due to the addition of an electron/hole to the state $\phi$.

Note that the $GW$ and $GW\Gamma_x$ self-energies can be written by Eqs.~\ref{sto_sigmaP_gw_deln} and \ref{sto_sigmaP_gwg_deln} by virtue of the stochastic decomposition of the GF. Only then, it is possible to express the correlation simply in terms of the induced time polarization and exchange potentials $W_P$ and $W_x$. 

\subsubsection{Stochastic induced time-dependent density and density matrix}\label{sssec:stochdenmtx}

In this part, we describe how we evaluate both $\delta{n}$ and $\delta{\rho}$ using \emph{stochastic sampling} rather than by summation over all occupied states. Since the density matrix is not constructed from eigenstates of $\hat{H}_0$, it will naturally fluctuate in time unless an infinite number of stochastic vectors is used. 

In practice, the density matrix induced by the addition/removal of an electron is computed as
\begin{equation}
\delta \rho\left(\br_1,\br_2,t_1\right) = \frac{\rho_\lambda(\br_1,\br_2,t_1) - \rho_{0}(\br_1,\br_2,t_1)}{\lambda}.\label{indstodensmat_t}
\end{equation}
Here, $\rho_{\lambda}$ represents the perturbed density matrix and $\lambda$ denotes the strength of the perturbing potential due to charge addition (see discussion below Eq.~\ref{deltan_response_pot}). $\rho_{0}$ is the unperturbed density matrix which exhibits time dependence due to its stochastic nature. 

The time-dependent density matrices are constructed from random vectors in the occupied subspace 
\begin{equation}
\rho_\lambda\left(\br_1,\br_2,t_1\right) = \left\{ \eta_\lambda \left(\br_1,t\right)  \eta_\lambda^* \left(\br_2,t\right) \right\}.\label{stodensmat_t}
\end{equation}
The stochastic states $\eta_\lambda (t)$ are found for each $t$ by time evolution according to:
\begin{equation}
\eta_\lambda(\br_1,t_1) =  \left\langle \br_1 \middle |e^{-i\hat{H}_0(t_1) t_1}\middle |\eta_{\lambda}\right\rangle, \label{tprop_eta}
\end{equation}
where $\hat{H}_0$ is the GKS Hamiltonian from Eq.~\ref{RSHHamiltonian}, which adiabatically depends on $t$ since $\hat\Sigma_H$, $\hat{V}_c$ and $\hat{V}_x$ are functionals of the time-dependent density, and $\hat\Sigma^\gamma_x$ is a functional fo the density matrix. The time-dependent density is, of course, $n(\br_1,t_1) = \rho(\br_1,\br_1,t_1)$. 
Numerically, Eq.~\ref{tprop_eta} is solved using Trotter propagation technique (see Sec.~\ref{sec:tprop_stoch_LRX}). 

The states $\eta_\lambda$, are perturbed at $t=0$:
\begin{equation}
\left| \eta_\lambda \right\rangle = e^{i\hat{v}_{\lambda}} \left| \eta \right\rangle,
\end{equation}
where $\hat{v}_{\lambda}$ is a perturbing potential:
\begin{equation}
v_\lambda(\br_1) = \lambda \nu(\br_1, \bar{\br}_2) \bar\zeta(\bar{\br}_2) \phi_i(\bar{\br}_2).
\end{equation}
Here, $\lambda$ is the strength of the perturbation. In practice, we take $\lambda = 10^{-4} E_h^{-1}$, but the value of $\lambda$ between $10^{-5} E_h^{-1}$ and $10^{-3} E_h^{-1}$ does not affect the results for molecules in Sec.\ref{sec:results}. This is consistent with previous observations.\cite{neuhauser2014breaking,rabani2015time,vlcek2017stochastic,vlvcek2018swift,vlcek2019stochastic}

In practice, stochastic computation of the induced charge density requires only a few random states $\eta$ (typically between 4 and 20), i.e., the number of states that are propagated by Eq.~\ref{tprop_eta} is much smaller than the number of occupied states. Further, the induced density matrix is damped by a factor $\exp[-(\alpha t)^2 / 2]$, where the damping factor is related to the maximum propagation time $\alpha = 3/t_{max}$.    

\subsubsection{Time propagation and stochastic decomposition of the long-range exchange -- RPA and TDDFT response}\label{sec:tprop_stoch_LRX}

The self-energy requires two distinct time-propagations to be computed: for the Green's function (Eq.~\ref{tprop_zeta}), and for the density matrix (Eq.~\ref{tprop_eta}). In both cases, the time-evolution operator is split into the local and non-local part of $\hat{H}_0$, and it is calculated in discrete time steps $\Delta t$ as:
\begin{equation}
e^{-i\hat{H}_0 \Delta t} = e^{-i\hat\Sigma_x^\gamma \frac{\Delta t}{2}} e^{-i\hat{h}_L \Delta t}e^{-i\hat\Sigma_x^\gamma \frac{\Delta t}{2}},
\end{equation}
where the local part of the Hamiltonian is
\begin{equation}
\hat{h}_L \equiv \hat{h} + \hat\Sigma_H  + \hat{V}_c + \hat{V}_x .
\end{equation}
The time-evolution due to the non-local part is computed simply as:
\begin{equation}
e^{-i\hat\Sigma_x^\gamma \frac{\Delta t}{2}} \approx \left( \hat{\mathcal{I}}- i\hat\Sigma_x^\gamma\frac{\Delta t}{2}\right).\label{tprop_NLX}
\end{equation}
Here, the time-step is a parameter subject to convergence tests; in typical calculations,  $\Delta t$ ranges between 0.02 and 0.05 a.u.

There are only a few vectors $\zeta$ and $\eta$. Nevertheless, the time evolution is costly due to the non-locality of $\Sigma^\gamma_x$. To make Eq.~\ref{tprop_NLX} less expensive, we use two additional sets of stochastic vectors to represent  $\Sigma^\gamma_x$ (Eq.~\ref{sigma_gx}):

(i) The first set is used for the long-range Coulomb interaction $\nu^\gamma$:\cite{neuhauser2015stochastic}
\begin{equation}
\nu^\gamma(1,2)  = \left\{\chi^\gamma\left(\br_1\right) \chi^\gamma\left(\br_2\right) \right\} \delta(t_1,t_2). \label{coulX_decomp}
\end{equation} 
This form was applied previously to the ground state calculations.\cite{neuhauser2015stochastic,vlcek2019stochastic} Here it is applied only to the time-evolution of stochastic states.

(ii) The second set, $\left\{\vartheta \right\}$, is used to decompose $G_0$. Since the exchange interaction is instantaneous, the GF is merely a density matrix.  In practice, it is sufficient to use one or two the stochastic states $\vartheta$ (see the next section), which are obtained by a linear combination:
\begin{equation}
\vartheta(1) = \frac{1}{N_\eta} \sum_{j=1}^{N_\eta} e^{i\theta_j(t_1)} \eta_j(1),\label{thetastates}
\end{equation}
where $\theta\in [0,2\pi]$ is a random phase, and $N_\eta$ is the number of $\eta$ vectors used for decomposition of the density matrix (Eq.~\ref{stodensmat}). 

Together, the action of $\hat\Sigma_x^\gamma$ on an arbitrary vector $\psi$ is
\begin{equation}
\left \langle \br_1 \middle |\hat{\Sigma}^\gamma_x \middle | \psi \right\rangle = \left\{ \vartheta(\br_1) \chi^\gamma\left(\br_1\right)\chi^\gamma\left(\bar{\br}_2\right) \vartheta\left(\bar{\br}_2\right) \psi(\bar\br_2) \right\}.\label{stoch_X_self}
\end{equation}
Time arguments are omitted here since the exchange interaction is instantaneous (note the delta function in Eq.~\ref{coulX_decomp}).  Also, note that the $\bar\br_2$ coordinates are integrated out, i.e.,  $\chi^\gamma\left(\bar{\br}_2\right) \vartheta\left(\bar{\br}_2\right) \psi(\bar\br_2) $ is a complex number. 

In practice, the numbers of $\vartheta$ and $\chi$ vectors are finite, and hence the stochastic noise is, in principle, increased.  However, at each time step, new random phases $\left\{\theta_j\right\}$  are selected. Frequent resampling of $\vartheta$ helps to reduce the stochastic error. As a result, only a few states are needed in actual calculations (see Sec.~\ref{sec:results}). 

Finally, it is necessary to point out the difference between response computed with random phase approximation (RPA) and TDDFT. In the first case, only the Hartree self-energy evolves in time (i.e., it is constructed at each time step).  Such treatment corresponds to the time-dependent Hartree approximation (equivalent to RPA). Here, the exchange and correlation terms are static. The non-local exchange interaction is computed with $\vartheta$ that repeatedly sample the \emph{static} unperturbed vectors $\eta$ at $t=0$. The corresponding results are labeled as $G_0W_0$. 

If the screening is computed with TDDFT, both Hartree and xc terms are constructed from the time-dependent states. The $\vartheta$ vectors that are made at each time step by a linear combination of time-evolved   $\eta_\lambda$ states.
 To distinguish the level of theory applied, we label the beyond-RPA approaches as $G_0W_0^{tc}$ and $G_0W_0^{tc}\Gamma_X$, because the screened Coulomb interaction is based on a test charge-test charge response function.\cite{bruneval2005many} All the methods are tested in the next section.
 
\section{Results}\label{sec:results}

\subsection{Verification of the time-dependent formulation}

In this section, we use the time-dependent formulation of the self-energy (derived in Sec.~\ref{subsec:sigma_approx}) combined with the stochastic approach described in Sec.~\ref{sec:methods}. We first compute the HOMO energies of small molecules using HF starting point. The results are verified against deterministic calculations in the frequency domain from Ref.~\citenum{maggio2017gw}.

We deliberately limit the stochastic approach to the decomposition of the Green's functions. Another set of stochastic orbitals is used only for sparse stochastic compression and time-ordering of the induced potentials.\cite{vlvcek2018swift} We use Eqs.~\ref{sto_sigmaP_gw_deln} and \ref{sto_sigmaP_gwg_deln}, in which $W_P(t)$ and $W_x(t)$ are computed deterministically. This partially stochastic approach is chosen because, for small systems, the stochastic time propagation leads to substantial statistical errors, which decrease only slowly with the number of stochastic states.\cite{vlcek2017stochastic,vlcek2019stochastic} 

The ground state electronic structure was computed on a real-space grid for selected molecules (Table~\ref{tab:HFbench}). In all cases, a grid of $64^3$ points with a spacing of $h=0.30~a_0$ yields  results converged to 0.02~eV. Our calculations are performed only for valence electrons; we use LDA Trouiller-Martins pseudopotentials\cite{troullier1991efficient} and kinetic energy cutoff of 28~$E_h$. 

As discussed in Sec.~\ref{ssec:stochastics}, the time propagations of $G_0$ and $\delta \rho$ are performed in discrete steps $\Delta t$ for a limited total simulation time $t_{max}$. 
We use a time-grid spacing of 0.05~a.u.;  $t_{max}$ was 100~a.u. This yields converged results with two exceptions: in $G_0W_0^{tc}$ and $G_0W_0^{tc}\Gamma_x$ runs for ethylene and methane, $t_{max}=150$~a.u.~was necessary due to a low-frequency features discussed later. Note that $\Delta t$ and $t_{max}$ are parameters of the calculation (similar to grid size and spacing), and hence they affect the statistical error only indirectly. The GF was sampled by $N_\zeta$ stochastic vectors (Table.~\ref{tab:HFbench}); the value of $N_\zeta$ was converged so that the statistical errors are less than 0.05~meV. $N_\zeta$ varies strongly among different systems and methods. 

The $G_0W_0$ results (Table.~\ref{tab:HFbench}) are in excellent agreement with the calculations from Ref.~\citenum{maggio2017gw} extrapolated to the complete basis-set limit. The mean absolute error (MAE) is 0.16~eV, which is a typical discrepancy between two distinct implementations (real-space and real-time versus atomic basis-set and frequency-domain).\cite{van2015gw} In all cases, $N_\zeta<5,000$, which is slightly lower compared to the fully stochastic $G_0W_0$ based on a KS DFT starting point.\cite{vlcek2017stochastic} 

Inclusion of vertex corrections shifts QPs to higher energies.  $G_0W_0^{tc}$ usually provides higher estimates than $G_0W_0^{tc}\Gamma_x$.
To facilitate the comparison between our results and Ref.~\citenum{maggio2017gw}, we only consider differences between the $G_0W_0$ and the other methods. Our predictions are in excellent agreement with previous calculations yielding MAE of 0.04 and 0.07 eV for $G_0W_0^{tc}$ and $G_0W_0^{tc}\Gamma_x$. 

\begin{figure}
\includegraphics[width=0.475\textwidth]{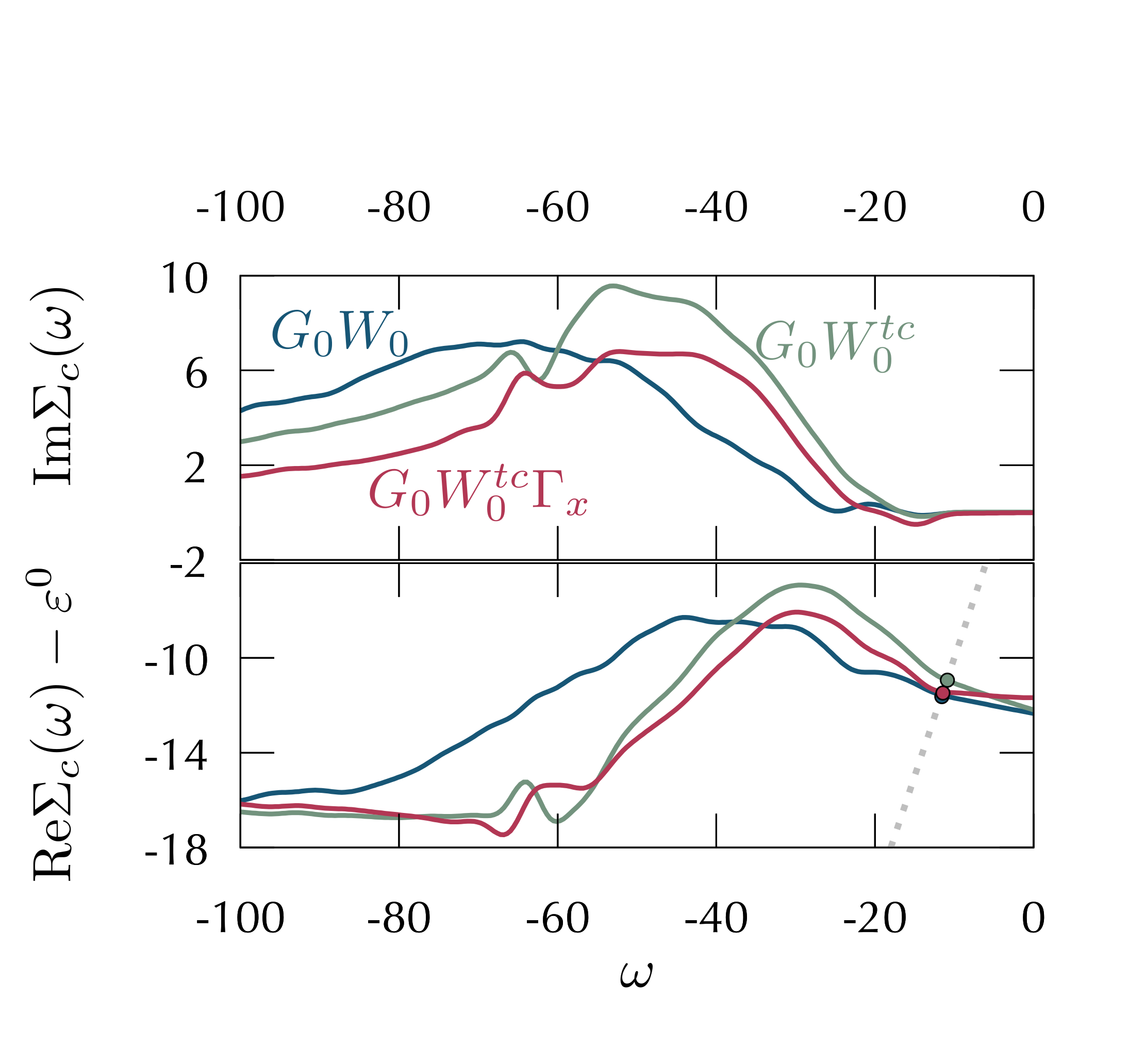}
\caption{The self-energy of methanol computed stochastically with three approximation to the correlation self-energy (distinguished by color and labeled in the graphs). The upper panel shows the imaginary part of the self-energy. The lower panel is the graphical solution to Eq.~\ref{QP_eq_lin_DFT}: the dashed gray line represents the frequency, and the intersections (marked by circles) correspond to the QP energies given in Table.~\ref{tab:HFbench}. The numbers of stochastic vectors are in  Table.~\ref{tab:HFbench}. All units are in eV. }\label{fig:meoh_se}
\end{figure}

Fig.~\ref{fig:meoh_se} illustrates the self-energy of methanol calculated with the three distinct methods using the stochastic sampling of the GF. The spectral features in the self-energy (both real and imaginary part) are broadened due to a finite length of the time propagation. The present frequency resolution is,  however, sufficient. In small systems, Eq.~\ref{QP_eq_lin_DFT} requires self-energy at frequencies sufficiently distant from the poles of $\Sigma_c$, where the curves are smooth and monotonic. We have increased the computational time by 50\% and found that the QP energies change by $<0.02$~eV.

The vertex corrections tend to shift the spectral features to lower frequencies, as shown in Fig.~\ref{fig:meoh_se}. Hence, the variation of the self-energy is extended over longer time-scales in $G_0W_0^{tc}$ and $G_0W_0^{tc}\Gamma_x$. This explains why longer propagation times were needed in some calculations which included vertex (namely for ethylene and methane). Further, the vertex function in the screened Coulomb interaction tends to increase the amplitude of the frequency variation of $\Sigma_c$. More stochastic samples are thus needed to converge the calculation to the same level of statistical error, which is also seen in the results in Table.~\ref{tab:HFbench}.

\begin{table*}
\begin{tabular}{c| c c c c c}
system	  & $G_0W_0$ ($N_\zeta$)	&$\Delta G_0W_0^{tc}$  ($N_\zeta$)&Ref.~$\Delta G_0W_0^{tc}$  &  $\Delta G_0W_0^{tc} \Gamma_X$  ($N_\zeta$)& Ref.~$\Delta G_0W_0^{tc} \Gamma_X$ \\
\hline
ammonia 			&-11.17 (4400)&0.56 (8800)&0.59&0.13 (4300) &0.17 \\
ethylene    		&-11.02 (1700)&0.17 (3200)&0.20&0.36 (1700) &0.25 \\
methane			&-15.04 (1500)&0.40 (2800)&0.40&0.45 (1400) &0.38 \\                                                                                                                
methanol  			&-11.56 (3500)&0.64 (6300)&0.61&0.28 (3000) &0.32 \\
water	  			&-13.25 (4800)&0.71 (9200)&0.80&0.26 (5100) &0.18 \\
\end{tabular}
\caption{Quasiparticle energies computed with distinct approximations to the self-energy based on HF starting point. Statistical error of each calculation is $\le0.05$~eV. The numbers of stochastic vectors representing the Green's function ($N_\zeta$) are given in the parentheses. $N_\zeta$ is rounded up to the nearest 100. $\Delta G_0W_0^{tc}$  represents the difference between the $G_0W_0^{tc}$ and  $G_0W_0$ result; the same notation is used for $G_0W_0^{tc} \Gamma_X$ in the last two columns. Reference results (labeled Ref.~in the column header) are taken from Ref.~\citenum{maggio2017gw}. All values are in eV. }\label{tab:HFbench}
\end{table*}

\subsection{Fully stochastic method}

We now turn to the calculations of the QP energies using a fully stochastic approach. Treating HF exchange by stochastic sampling may require a very high number of stochastic vectors leading to a high computational cost. However, it is possible to decompose the long-range non-local exchange interactions in GKS DFT.\cite{neuhauser2015stochastic,rabani2015time,vlcek2019stochastic} 

Here, we apply the LC-$\omega$PBE functional discussed in Sec.~\ref{ssec:RSH_DFT}. Both the GF and the time-dependent potentials ($W_P$ and $W_x$ in Eqs.~\ref{sto_sigmaP_gw_deln} and \ref{sto_sigmaP_gwg_deln})  are sampled stochastically using multiple sets of stochastic vectors. 

We calculate HOMO and LUMO QP energies of  the molecules listed in Table~\ref{tab:mol_grids}. The results were converged with respect to the real-space grid; the number of grid points ($N_g$) is specified in Table.~\ref{tab:mol_grids} for each system. In all calculations, we used the grid spacing of $h=0.35~a_0$. The ground state eigenvalues are converged with respect to the grid parameters to $<0.01$~eV. The QP energies are converged to within 0.03~eV. Fig.~\ref{fig_converge} shows that the differences in the self-energy for grids with $h=0.35$ and $h=0.30~a_0$ are small. 

The range-separation parameter ($\gamma$) was selected such that the ionization potential theorem of the neutral system is satisfied. Tuning $\gamma$ to enforce the ionization potential simultaneously for the neutral molecule and an anion tends to change the parameter negligibly. Indeed, the DFT results presented here are in agreement with Ref.~\citenum{rangel2016evaluating} , in which the latter tuning approach was applied. The mean absolute deviation from Ref.\citenum{rangel2016evaluating}  is 0.10/0.06~eV for HOMO/LUMO.

We focus only on relatively large molecules that form stable anions since the goal is to test how different approximations treat both ionization potential and electron affinities. Further, we investigate how the computational cost scales with the system size. The convergence of the computed QP energies is described in the next section; the results of the three methods are compared afterward.

\begin{table}
\caption{Real-space grids (characterized by the number of points $N_g$) and the range-separation parameters ($\gamma$) used in the calculations. The molecular structures were taken from Refs.~\citenum{dorset1994disorder,rangel2016evaluating,casalegno2013solvent} }\label{tab:mol_grids}
\begin{tabular}{c| c c}
system  &$N_g$ &$\gamma$ [$a_0^{-1}$]\\
\hline
anthracene 			& $80 \times  60\times 50$ &0.23    \\
tetracene    			& $88   \times60 \times 50$ &0.21    \\
pentacene  			& $108 \times60\times 50$ &0.19    \\
hexacene   			& $112 \times60\times 50$ &0.17    \\
C$_{60}$     			& $88  \times 88\times 88$ &0.18    \\
PC$_{60}$BM   	& $88 \times  88\times 88$ &0.15    \\                                                    
\end{tabular}

\end{table}

\subsubsection{Convergence of stochastic errors}

\begin{figure}
\includegraphics[width=0.475\textwidth]{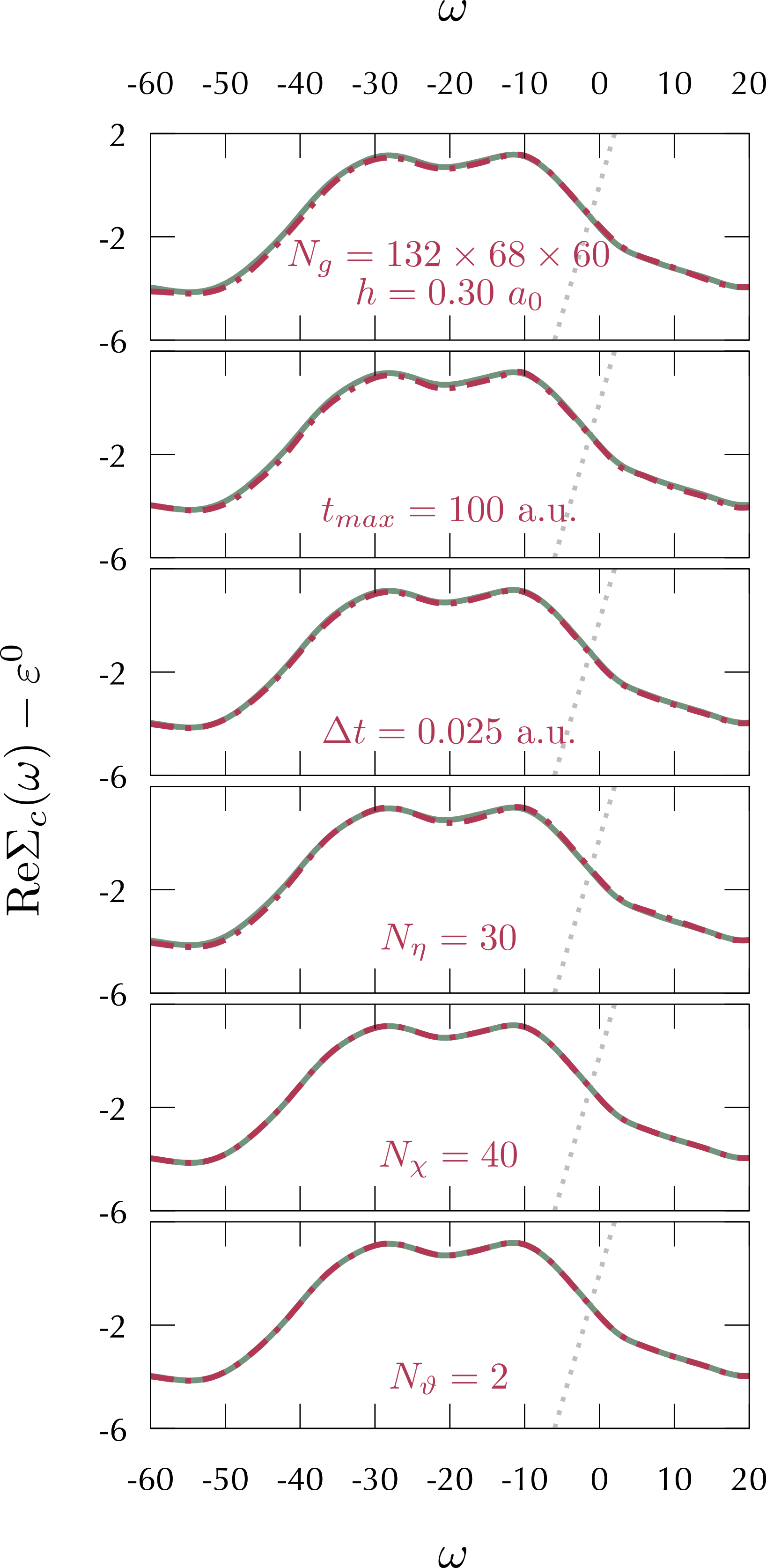}
\caption{The self-energy of LUMO in hexacene computed with stochastic $G_0W_0^{tc}\Gamma_x$ method with varying parameters of the grid, time propagation length, time step, and the numbers of stochastic states $N_\eta$, $N_\chi$ and $N_\vartheta$ in each panel from top to bottom. The panels show the graphical solutions to Eq.~\ref{QP_eq_lin_DFT}: the dashed gray line represents the frequency, and the intersections correspond to the QP energies. The full green lines represent the solution obtained with $N_g = 112 \times 60 \times 50$, $h= 0.35 a_0$, $t_{max}=50$a.u., $N_\eta=15$, $N_\chi=20$ and $N_\vartheta=1$. The red dash-dotted line represents a solution where one of the parameters is changed as labeled in the graph.  All units are in eV. }\label{fig_converge}
\end{figure}

In the fully stochastic implementation, the statistical errors arise predominantly from the induced time-dependent potentials $W_P$ and $W_x$ in Eqs.~\ref{sto_sigmaP_gw_deln} and \ref{sto_sigmaP_gwg_deln}. The time evolution is performed with $\Delta t=0.05$~a.u.~and $t_{max} = 50$~a.u. The propagation time is shorter than for small molecules with HF starting point because the dynamics with LC-$\omega$PBE functional exhibits a faster time decay of the response. These values of $\Delta t$ and $t_{max}$ yield QP energies converged to better than $0.02$~eV. This is consistent with the previous stochastic calculations for $G_0W_0$ based on LDA starting point.\cite{vlcek2017stochastic,vlvcek2018swift}  Increasing $t_{max}$ or decreasing $\Delta t$  affects neither the QP energies nor the self-energies, as illustrated in Fig.~\ref{fig_converge}.

For a given set of time and real-space grid parameters, the QP energies exhibit stochastic fluctuations stemming from the random sampling vectors. In all three approximations, the following sets of stochastic orbitals are employed: (i) $\zeta$ for decomposition of the GKS Green's function (Eq.~\ref{Gtexpansion}); (ii) $\eta$ for decomposition of the induced density matrix (Eq.~\ref{stodensmat_t}); (iii) $\chi$ for the decomposition of the exchange kernel $\nu^\gamma$ (Eq.~\ref{coulX_decomp}); (iv) $\vartheta$ for decomposition of the density matrix in $\Sigma_x^\gamma$ (Eq.~\ref{thetastates}). 

Three types of stochastic vectors are part of a ``nested sampling'':  There are $N_\eta \times N_\chi\times N_\vartheta$  states per each $\zeta$ vector. The overall error is thus governed mainly by the number of $N_\zeta$ samples, each having a stochastic fluctuation determined by $N_\eta \times N_\chi\times N_\vartheta$. In practice, $N_\zeta$ is increased until the statistical error is below a predetermined threshold.

For all systems investigated, we take $N_\eta=15$, which is consistent with previous calculations for acenes and C$_{60}$.\cite{vlvcek2018swift}  Additional tests for anthracene and C$_{60}$ show that the same QP energies are obtained with $N_\eta = 12$ (albeit with higher stochastic fluctuation per single $\zeta$). Fig~\ref{fig_converge} illustrates that doubling the number of stochastic vectors $\eta$, i.e., $N_\eta=30$, does not affect the results. 

For the other two sampling vectors, it is sufficient to take $N_\vartheta=1$ and $N_\chi=20$. Such low values are due to a small magnitude of the $\Sigma_x^\gamma$ term, which stems from a weak long-range exchange ( $\gamma \le 0.23\,a_0^{-1}$, see Table.~\ref{tab:mol_grids}).  Fig.~\ref{fig_converge} shows that twice as high numbers of stochastic vectors $N_\vartheta$ and $N_\chi$ again does not change the results. The self-energy curves are almost identical; they differ by $<0.02$~eV at the QP energy, which is much less than the statistical error due to finite $N_\zeta$ and $N_\eta$. 

The overall error of the fully stochastic approach accumulates each of the contributions discussed above (i.e., real-space and time grids, and the numbers of stochastic vectors). Calculations for tetracene and  hexacene LUMO  states showed that the total (accumulated) error is $<0.04$~eV. The error was estimated by comparing the results in Table~\ref{tab:QP_ene} with results for $h=0.30$, $\Delta t=0.03$~a.u., $t_{max}=100$~a.u., $N_\eta = 30$, $N_\vartheta=2$ and $N_\chi=40$. The error is relatively low due to mutual cancellation among the different contributions. This is consistent with previous calculations for molecules.\cite{vlcek2017stochastic,vlvcek2018swift}

\begin{figure}
\includegraphics[width=0.5\textwidth]{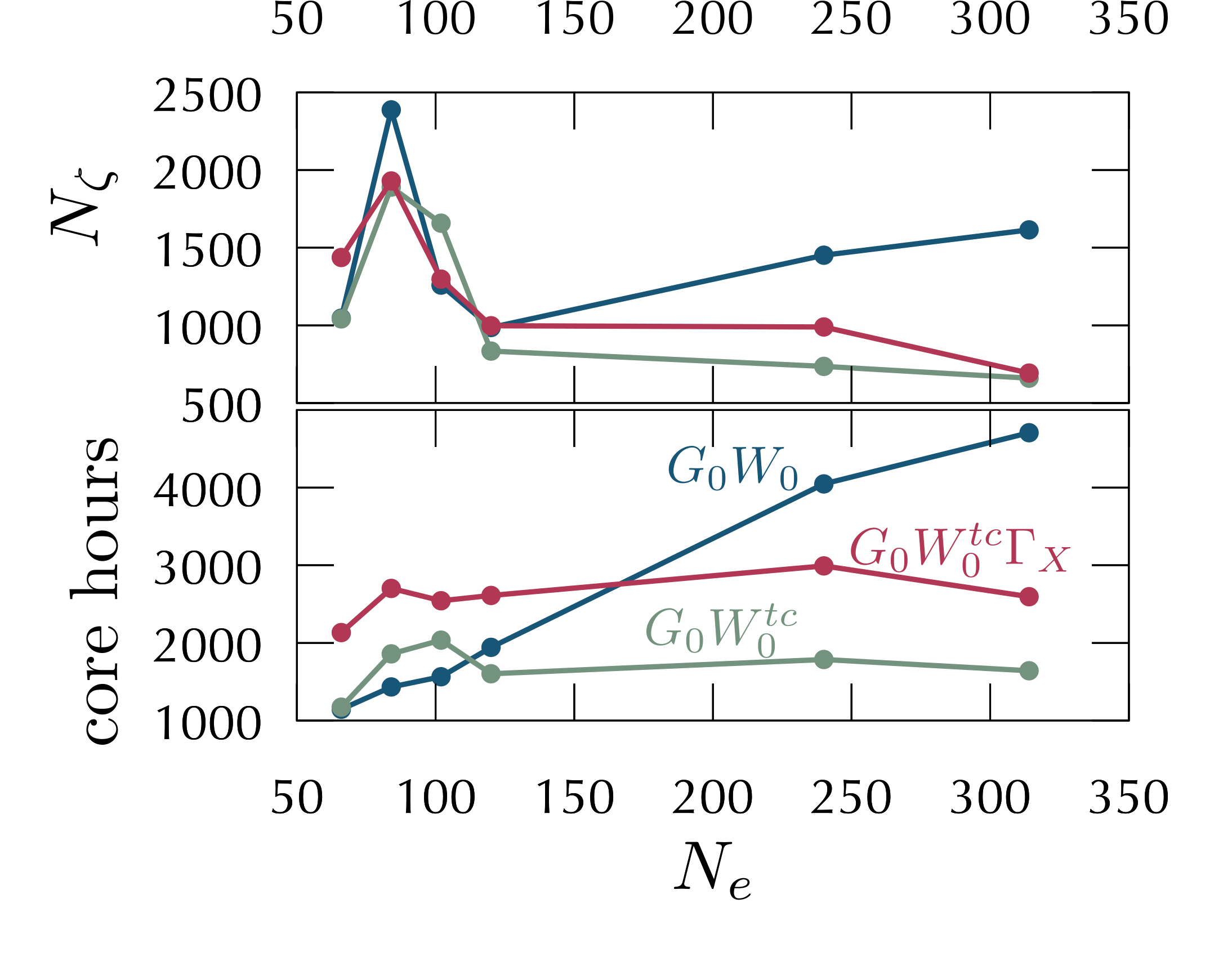}
\caption{The top panel shows the number of $\zeta$ states required to converge the statistical error of the HOMO and LUMO QP energies below 0.05~eV as a function of the total number of valence electrons $N_e$. The bottom graph shows the total number of core hours required for calculations of the QP energies. The calculations were performed on Bridges computer equipped with Intel Haswell (E5-2695 v3) CPUs. }\label{times_samples}
\end{figure}

Finally, we compare the total stochastic error in the different approximations to $\Sigma_c$. In this analysis, the target fluctuation, $\sigma(\varepsilon)$, is 0.05~eV. The results are shown in Fig.~\ref{times_samples}.   
For large systems,  the  $G_0W_0$ calculations converge slower compared to the other approximations, yet the computational cost maintains linear scaling (with a steep slope of $\sim 15$ core hours per electron).  Further, $N_\zeta$ rises with system size for the two largest molecules. In contrast, the costs of $G_0W_0^{tc}$ and $G_0W_0^{tc}\Gamma_X$ depend much less on the  system size and their computational time remains practically constant for systems between 100 and 300 valence electrons.

\begin{figure}
\includegraphics[width=0.45\textwidth]{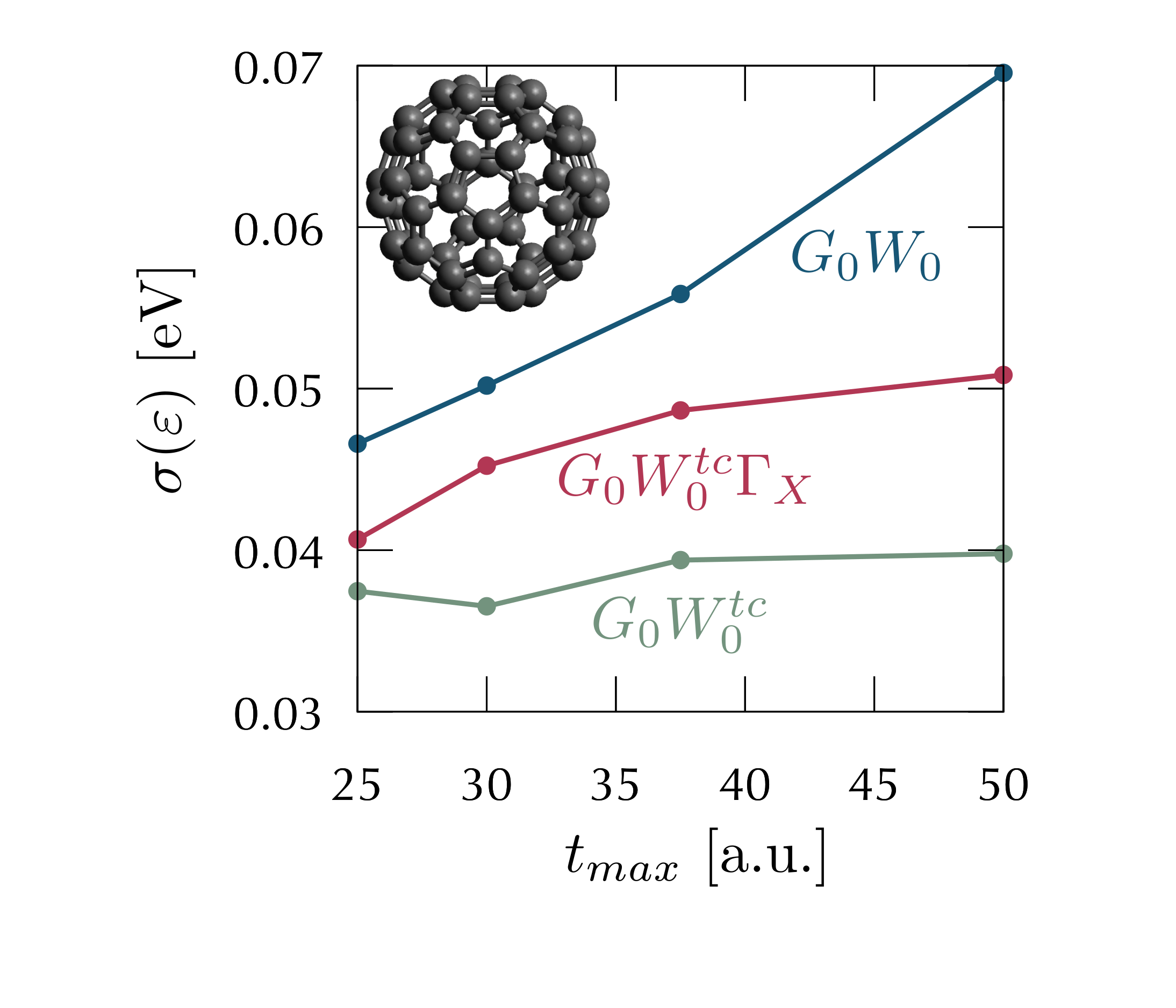}
\caption{The statistical error, $\sigma(\varepsilon)$, of the self-energy is shown as a function of the maximum propagation time, $t_{max}$, used for the calculation of the polarization self-energy. $\sigma(\varepsilon)$ is evaluated at the frequency corresponding to the QP energy $\varepsilon$. The data are for the LUMO QP state of the C$_{60}$ molecule. The number of $\zeta$ states is 1000, and the remaining parameters are described in the text.  }\label{tmax_RPA}
\end{figure}

The distinct behavior of the stochastic $G_0W_0$ calculations is due to RPA applied in the time propagation. As discussed in Sec.~\ref{sec:tprop_stoch_LRX}, RPA assumes that $\Sigma_x^\gamma$ is time-independent, i.e., it is not constructed from time propagated states. Although the $\Sigma_x^\gamma$ term is sampled by distinct stochastic vectors $\vartheta$ at each time step, it leads to a strong stochastic noise. This random fluctuation is amplified with time (similar to the breakdown of stochastic BSE\cite{rabani2015time}). Tests for hexacene showed that taking $N_\vartheta=2$ leads to only a $\sim 1\%$ reduction of the fluctuation. Fig.~\ref{tmax_RPA} clearly illustrates the amplification of the stochastic error in $G_0W_0$ with $t_{max}$. In contrast, $\sigma(\varepsilon)$ in $G_0W_0^{tc}$ calculation remains almost constant regardless of $t_{max}$. In $G_0W_0^{tc}\Gamma_X$ calculations, the statistical error is higher, but it does not increase significantly with  $t_{max}$.

In general, the stochastic approach is aimed for large systems.\cite{baer2013self,neuhauser2014breaking,vlcek2017stochastic,vlvcek2018swift} The calculation for tetracene requires the highest number of stochastic samples irrespective of the method chosen (Fig.~\ref{times_samples}). This behavior can be understood as follows: For systems smaller than tetracene, $N_\eta=15$ is relatively high compared to the number of occupied states (for instance, there are 33 occupied valence states in anthracene). As a result, the occupied subspace is sufficiently well sampled. For large systems, the stochastic approach exhibits strong self-averaging,\cite{baer2013self,neuhauser2014breaking,vlvcek2018swift} which leads to a decrease of $N_\zeta$  required for target $\sigma(\varepsilon)$, i.e., the computational cost decreases. Tetracene is found to be the ``worst-case scenario'' in which the stochastic sampling introduces relatively large errors, and there is only limited self-averaging.

Overall, the fully stochastic implementation of $G_0W_0^{tc}$ and $G_0W_0^{tc}\Gamma_X$ is efficient and numerically stable, while $G_0W_0$ suffers from stochastic fluctuations. The total computational time of the beyond-RPA-methods depends only weakly on the system size  (Fig.~\ref{times_samples}).  For large systems, the more involved expressions for the self-energy are less expensive than their $G_0W_0$ counterpart. The low cost of stochastic beyond-$GW$ calculations is in striking contrast to their conventional (deterministic) implementations.

\subsubsection{Performance of the approximations to $\Sigma_c$}

We will now address how distinct approximations to $\Sigma_c$ affect predictions of the HOMO and LUMO energies. Here, we will report results of stochastic calculations with $N_\zeta=1500$,  $N_\vartheta=1$, $N_\chi = 20$, and $N_\eta=15$. The DFT starting point (optimally tuned LC-$\omega$PBE) is already in a good agreement with the reference values, as shown in Table~\ref{tab:QP_ene}; yet, the HOMO/LUMO energies are consistently over/underestimated. The mean absolute error (MAE) of the DFT reference point for acenes is 0.17~eV for the HOMO energies and 0.28~eV for the LUMO energies.

The $G_0W_0$ approximation makes the HOMO energies more negative (Table~\ref{tab:QP_ene}). As a result, the ionization potentials (negative of the HOMO energy) are significantly improved, and the corresponding MAE is only 0.07~eV. The $G_0W_0$ performance is, however, very different for LUMO. Here, the correction is too large, and the QP energies are thus much higher than the reference values leading to MAE of 0.28 eV.  If we exclude the experimental reference data for C$_{60}$ and PC$_{60}$BM, we get MAE of 0.04~eV and 0.39~eV for HOMO and LUMO, which are in agreement with an earlier benchmark for acenes.\cite{rangel2016evaluating}

This is a disappointing result, because a more advanced computational technique ($G_0W_0$), which is aimed to improve upon DFT, yields worse results than the DFT itself. Further, the stochastic implementation of $G_0W_0$ on top of hybrid functionals is numerically expensive for large systems due to numerical instabilities discussed earlier. Note that such instabilities were not observed in previous stochastic $G_0W_0$ calculations based on LDA starting point.
 
The $G_0W_0^{tc}$ approximation is computationally stable and, similar to $G_0W_0$, it provides good ionization potentials; MAE for HOMO is 0.08~eV. In the systems selected, the presence of the vertex corrections thus does not have a pronounced effect on the occupied states. However, the method amplifies the problems for unoccupied states.  The affinities (negative of the LUMO energy) are predicted to be significantly larger than the reference values, leading to MAE of 0.44~eV which is worse than in $G_0W_0$. If experimental data for C$_{60}$ and PC$_{60}$BM are excluded from the analysis,  $G_0W_0^{tc}$ leads to even larger errors for LUMO (0.55~eV). This failure for unoccupied states indicates that abandoning RPA  has detrimental effects on unoccupied QP states.

\begin{figure}
\includegraphics[width=0.475\textwidth]{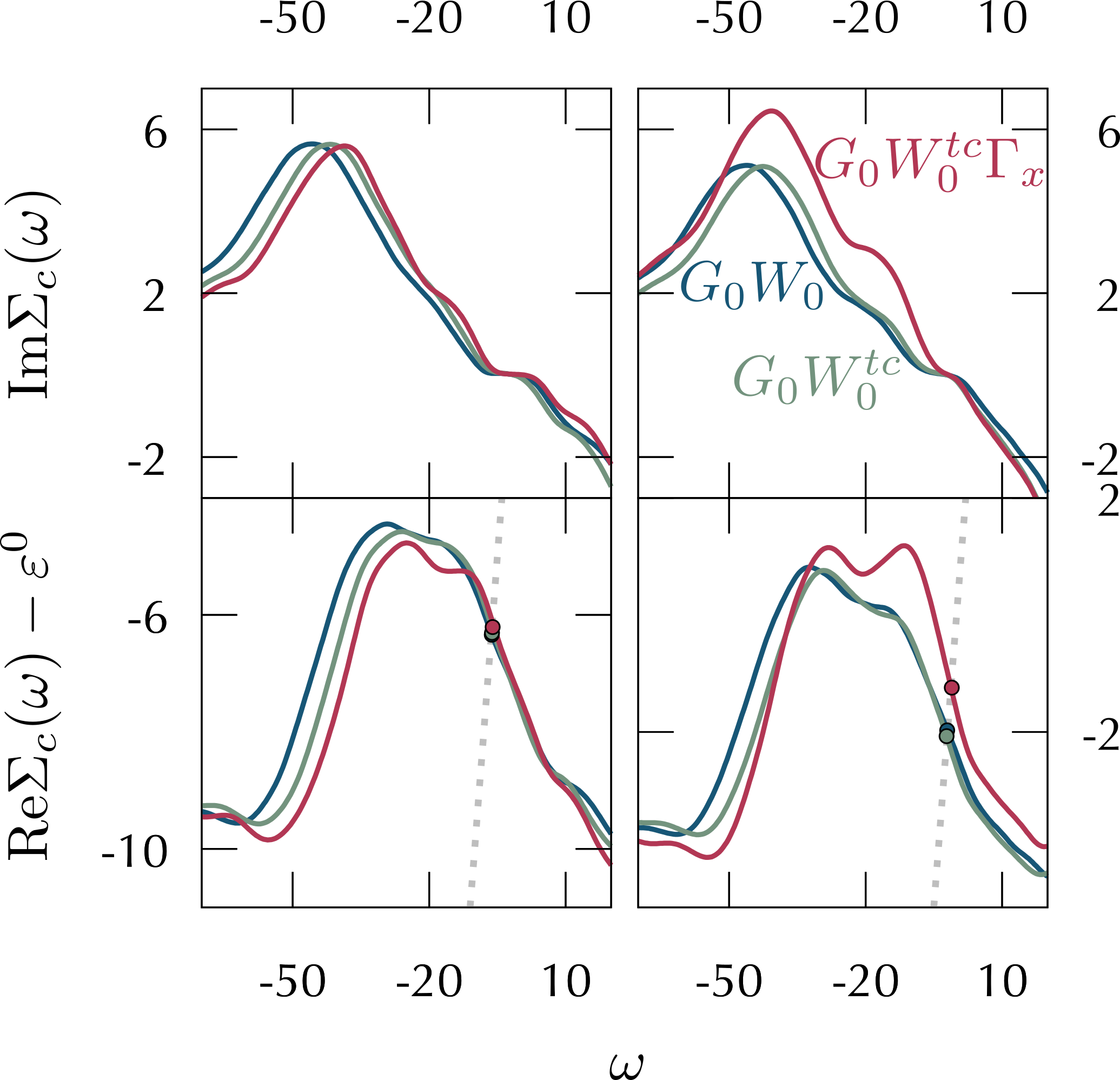}
\caption{The self-energy of hexacene HOMO and LUMO states (left and right panel) computed stochastically with three approximation to the correlation self-energy (distinguished by color and labeled in the graphs). The upper panel shows the imaginary part of the self-energy. The lower panel is the graphical solution to Eq.~\ref{QP_eq_lin_DFT}: the dashed gray line represents the frequency, and the intersections (marked by circles) correspond to the QP energies given in Table.~\ref{tab:QP_ene}.  All units are in eV. }\label{fig:hex_methods}
\end{figure}

Finally, we turn to the analysis of $G_0W_0^{tc}\Gamma_X$ predictions. 
The presence of the non-local exchange interaction in Eq.~\ref{DSt_dG_GWG} has a significant impact on the QP energies. The HOMO states are higher in energy, leading to MAE of 0.19~eV for acenes. This error is significantly worse than $G_0W_0$ but similar to the DFT results. 

In contrast to the other approximations tested here, $G_0W_0^{tc}\Gamma_X$ self-energies are of the unoccupied states are qualitatively different. Fig.~\ref{fig:hex_methods} illustrates that inclusion of the vertex changes the self-energy dramatically for the LUMO state (while for HOMO it is similar to the $G_0W_0$ and $G_0W_0^{tc}$ results). This observation is consistent with previous theoretical results based on local vertex corrections that incorporate derivative discontinuity.\cite{hellgren2018beyond,hellgren2018local} Here such discontinuity is included via the time-dependent induced exchange potential $W_x$ in the  $G_0W_0^{tc}\Gamma_X$  self-energy (Eq.~\ref{sto_sigmaP_gwg_deln}). The vertex correction thus acts similar to exchange in the static ground-state calculations and shifts unoccupied states higher in energy. In all cases tested, the QP energy of the unoccupied states are significantly increased and are higher than the reference values. Note that all other methods, the LUMO QP energies are too low.  For the acenes molecules, we find MAE of 0.26~eV; hence, the perturbation theory slightly improves upon the DFT starting point. 

Based on the results for acenes LUMO energies, $G_0W_0^{tc}\Gamma_X$ appears to be more successful than self-consistent $GW$ methods benchmarked in Ref.~\citenum{rangel2016evaluating}. If eigenvalue self-consistent $GW$ is employed and both the GF and the screened Coulomb interactions are updated, the predictions yield MAE of 0.41~eV. For the same subset of systems,  $G_0W_0^{tc}\Gamma_X$ yields MAE that is  $\sim50\%$ smaller. In an alternative scheme, the self-consistency is applied only to the GF; however,  for LUMO states the MAE increases to 0.44~eV, i.e., the deviation is again higher than for $G_0W_0^{tc}\Gamma_X$ results.

\begin{table*}
\begin{tabular}{c| c c c c c}
& \multicolumn{5}{|c}{HOMO}\\
system	  & DFT	&$G_0W_0$	&$G_0W_0^{tc}$ & $G_0W_0^{tc} \Gamma_X$	& Ref.\\
\hline
anthracene  			 &-7.33  &-7.42 ($\pm$0.04)  &-7.31 ($\pm$0.04)  &-7.25 ($\pm$0.05)  &-7.48  \\
tetracene   			 &-6.70  &-7.00 ($\pm$0.06)  &-6.89 ($\pm$0.05)  &-6.79 ($\pm$0.06)  &-6.96  \\
pentacene       		 &-6.47  &-6.65 ($\pm$0.04)  &-6.55 ($\pm$0.05)  &-6.42 ($\pm$0.05)  &-6.58  \\
hexacene    			 &-6.15  &-6.32 ($\pm$0.05)  &-6.22 ($\pm$0.04)  &-6.11 ($\pm$0.05)  &-6.32  \\
C$_{60}$         		 &-7.90  &-7.69 ($\pm$0.06)  &-7.68 ($\pm$0.04)  &-7.60 ($\pm$0.05)  &-7.69* \\
PC$_{60}$BM        &-7.27  &-7.42 ($\pm$0.06)  &-7.26 ($\pm$0.03)  &-7.20 ($\pm$0.04)  &-7.17* \\    
\hline
& \multicolumn{5}{|c}{LUMO}\\
system	  & DFT	&$G_0W_0$	&$G_0W_0^{tc}$ & $G_0W_0^{tc} \Gamma_X$	& Ref.\\
\hline
anthracene  			 &-0.49  &-0.54 ($\pm$0.04)  &-0.71 ($\pm$0.04)  &+0.00 ($\pm$0.05)  &-0.28\\
tetracene   			 &-1.04  &-1.22 ($\pm$0.06)  &-1.39 ($\pm$0.05)  &-0.55 ($\pm$0.06)  &-0.82\\
pentacene       		 &-1.53  &-1.65 ($\pm$0.05)  &-1.80 ($\pm$0.05)  &-0.91 ($\pm$0.05)  &-1.21\\
hexacene    			 &-1.84  &-1.94 ($\pm$0.05)  &-2.07 ($\pm$0.04)  &-1.27 ($\pm$0.05)  &-1.47\\
C$_{60}$         		  &-2.47  &-2.77 ($\pm$0.07)  &-2.91 ($\pm$0.04)  &-1.90 ($\pm$0.05)  &-2.68*\\
PC$_{60}$BM        &-2.46  &-2.61 ($\pm$0.07)  &-2.85 ($\pm$0.04)  &-1.95 ($\pm$0.04)  &-2.63*\\   
                                                                                                                                             
\end{tabular}
\caption{Quasiparticle energies computed with DFT (LC-$\omega$PBE) and distinct approximations to the self-energy. Statistical errors of the stochastic methods are given in the parentheses. The reference values are taken from Refs.~\citenum{rangel2016evaluating,C60EA,sun2016ionization,akaike2008ultraviolet,larson2013electron}.  The number of $\zeta$ states  is 1,500; the remaining parameters are described in the text.}\label{tab:QP_ene}
\end{table*}

\section{Conclusions and Outlook}\label{sec:conclusions}

In this work, we presented an efficient way for improving predictions of QP energies beyond the popular $GW$ approach using stochastic paradigm. In practice, this improvement amounts to the inclusion of nontrivial parts of the vertex function in the correlation self-energy ($\Sigma_c$). Here, an approximate non-local vertex correction was  derived from the functional derivative of the exchange self-energy. In principle, the self-energy should be found by a self-consistent set of expressions. In practice,  we employ only a one-shot correction scheme on top of a mean-field starting point, which includes a non-local long-range exchange. This approach is labeled $G_0W_0^{tc}\Gamma_X$, and it is compared to (stochastic) $G_0W_0^{tc}$ and $G_0W_0$ methods. 

A new stochastic formulation reduces the overall computational cost considerably. In contrast to the previous implementation of stochastic $G_0W_0$, it is possible to use Hartree-Fock and generalized Kohn-Sham starting points. 

The real-time formulation of all three methods was verified against previous results for small molecules computed in the frequency domain. Note that the time-domain version of $G_0W_0^{tc}\Gamma_X$ is entirely new. The Hartree-Fock starting point is used together with a stochastic sampling of the Green's function. The calculations are in excellent agreement with the reference values. 

For larger systems, we present a fully stochastic approach in which two additional sets of stochastic vectors sample the non-local vertex. For the large molecules investigated, we employed DFT with a long-range non-local exchange interaction. Such functional form is efficiently sampled even with a small number of stochastic vectors.  The, otherwise extremely involved, beyond $GW$ calculations can thus be performed for large molecules at a low cost. While deterministic implementations scale as $N_e^4$ for $GW$ and $N_e^6$ for $GW\Gamma$ (where $N_e$ is the number of electrons), the stochastic formulation scales (sub)linearly. In fact, we found that RPA is numerically unstable; its statistical error worsens with the system size and the simulation time. In contrast, more difficult $G_0W_0^{tc}$ and $G_0W_0^{tc}\Gamma_X$ calculations are stable and, paradoxically, computationally less expensive than $G_0W_0$.

The three stochastic approximations were tested on a set of acene molecules, C$_{60}$ and PC$_{60}$BM. The computational costs of $G_0W_0^{tc}$ and $G_0W_0^{tc}\Gamma_X$ were practically constant with the system size. The overall computational time required to converge  $G_0W_0^{tc}\Gamma_X$ QP energies was on average $\sim50\%$ higher than in $G_0W_0^{tc}$ due to increased statistical fluctuation.  The overall sublinear scaling is due to the rapid convergence of the statistical errors with the number of electrons. Hence, stochastic algorithms will be a method of choice for demanding beyond-$GW$ calculations or, at least, for efficient implementation of non-local vertex functions.

While  DFT with optimally-tuned range-separated hybrid functionals (LC-$\omega$PBE) provides a good starting point, some deviation from reference data is observed. One-shot $G_0W_0$ and $G_0W^{tc}_0$ improve the description of the ionization potentials compared to LC-$\omega$PBE, but severely increase errors for electron affinities. On average, both methods perform worse than DFT; the worst performance is observed for the $G_0W^{tc}_0$ approach.  

The $G_0W_0^{tc}\Gamma_X$ performs worse than DFT for the occupied states, but it improves the description of unoccupied states. While all other approaches underestimate the LUMO QP energies,  $G_0W_0^{tc}\Gamma_X$ predicts them to be higher than the reference values. The energy increase of the unoccupied QP states is due to the vertex correction based on time-dependent induced non-local exchange potential. For the set of molecules considered, $G_0W_0^{tc}\Gamma_X$  is the only method that outperforms DFT for the electron affinities. 

Previous calculations which included approximate vertex corrections were mostly applied to ionization potentials of small molecules that do not form stable anions. The accuracy of predicted electron affinities is another major and more sensitive indicator for performance assessment. 

Together, these findings indicate that beyond $GW$ schemes are crucial for an improved description of QP energies. As shown here, stochastic techniques make such calculations affordable even for large systems. Future steps are directed toward the formulation of better self-energy expressions that include higher-order interactions (beyond $G_0W_0^{tc}\Gamma_X$) to improve the prediction of QP energies further. In particular, the vertex terms stemming from the time-dependent induced correlation potential will be studied. Investigations of charge transfer systems with strong electron hole-interactions are underway.

\begin{acknowledgement}
 The calculations were performed as part of the XSEDE\cite{XSEDE} Project No. TG-CHE180051. Use was made of computational facilities purchased with funds from the National Science Foundation (CNS-1725797) and administered by the Center for Scientific Computing (CSC). The CSC is supported by the California NanoSystems Institute and the Materials Research Science and Engineering Center (MRSEC; NSF DMR 1720256) at UC Santa Barbara.
\end{acknowledgement}

\bibliography{lib_exc_gw}

\end{document}